\documentclass{article}
\usepackage{graphicx}
\usepackage{amsmath}
\usepackage{amssymb}
\usepackage{bm}
\usepackage{cite}

\begin{document}



\begin{center}
{\huge\textbf{Solution of the quantum finite square well problem
 using the Lambert W function}}\\
--------------------------------------------------------------------\\
Ken Roberts\footnote{Physics and Astronomy Department, 
Western University, London, Canada, krobe8@uwo.ca},
S. R. Valluri\footnote{Physics and Astronomy,
and Applied Mathematics Departments,
Western University, London, Canada;
King's University College, London, Canada}\\
March 26, 2014
\end{center}

\begin{abstract}
We present a solution of the quantum mechanics problem
of the allowable energy levels of a bound particle in a
one-dimensional finite square well.
The method is a geometric-analytic technique utilizing
the conformal mapping $w \to z = w e^w$ between two 
complex domains.
The solution of the finite square well problem can be seen
to be described by the images of simple geometric shapes,
lines and circles, under this map and its inverse image.
The technique can also be described using the Lambert W
function.  One can work in either of the complex domains,
thereby obtaining additional insight into the finite square
well problem and its bound energy states.  There are many
opportunities to follow up, and we present the method
in a pedagogical manner to stimulate further research in
this and related avenues.
\end{abstract}





\section{Introduction}
\label{sect-intro}

Quantum well models are important for the design of 
semiconductor devices, such as the quantum well 
infrared photodetector (QWIP) which is used
for infrared imaging applications;
see Schneider and Liu \cite{Schneider-2007} for an overview.
The QWIP relies upon a quantum well which has been sized so
that the energy of an electron in the first excited state 
is quite near the threshold of confinement in the well.
The QWIP is therefore very sensitive to the arrival of a 
single photon.  
There are many other uses of quantum well models in
nanostructures; the textbook by Harrison \cite{Harrison-2009}
is a good survey of the field.

The one-dimensional quantum finite square well (FSW) model is 
a familiar topic in most introductory quantum mechanics books;
see for instance Bransden and Joachain \cite{Bransden-1989}, section 4.6.  
After deriving a pair of equations to describe the bound energy
levels within the well, the solution is carried out by
graphical or computational methods.
It is sometimes said that the FSW problem does not have an 
exact solution, but there are in fact exact solutions
as presented in the papers of Burniston and Siewert
\cite{Burniston-1973, Siewert-1973, Siewert-1978} and 
others \cite{Paul-2000, Aronstein-2000, Blumel-2005,
Anastasselou-1984, Anastasselou-1985, Ioakimidis-1988}.
Those exact methods generally rely upon contour integration
in the complex plane.  

We have found a simple geometric method which can be used 
to describe the solutions of the one-dimensional
finite square well problem.
Our approach is analytic, using complex variables,
but does not rely upon contour integration as such.
We instead make a strong appeal to geometric imagination.
We focus on the description of the solution set via 
conformal mapping of simple geometric shapes
(lines and circles) between two complex domains, using
the mapping given by $w \to z = w \, e^w$.  Thus
our solution might be described as ``geometric-analytic".

In the rest of this introductory section we will describe
the motivation of this note.  In section \ref{sect-solution}
we will show the mathematical details of the FSW solution.
That solution will be presented at a gradual pace, to
introduce the technique and so that it may also be used for
teaching if desired.

\textbf{Motivation}

When solving the one-dimensional FSW problem, after
some initial definitions and discussion, a textbook
will arrive at the task of finding solutions
to one of the two equations
\begin{equation}
\label{equtanv}
  v \, \mathrm{tan} \, v = u
\end{equation}
or
\begin{equation}
\label{equcotv}
  v \, \mathrm{cot} \, v = -u
\end{equation}
together with the constraint
\begin{equation}
\label{equ2v2}
  u^2 + v^2 = R^2.
\end{equation}
Here $u$ and $v$ are positive reals
which are related to the allowed bound energy levels
which are to be found.
$R$ is a unitless parameter of the problem
which is determined only by the dimensions of
the potential well --
that is, its (spatial) width and (potential) depth --
and independent of the energy level of the bound state.
Bransden and Joachain call the parameter $R$
the ``strength parameter" of the problem, and
we will also use that terminology.

One is often shown graphical solutions of the 
pair of simultaneous equations
(\ref{equtanv}) and (\ref{equ2v2}),
or of the pair
(\ref{equcotv}) and (\ref{equ2v2}),
and is encouraged to use a computer
to find numerical solutions.
The allowable energy levels for the
bound particle are then calculated
from the values of $u$ or $v$.

In this note, we will show that the
solutions of the FSW problem can be 
described using the Lambert W function
\cite{Corless-1996, NIST-2010, Valluri-2000}.
This alternative solution may provide
some insight into the FSW problem.
Sometimes having a solution determined
by an analytic function, 
instead of graphically or numerically,
can make it easier to use that solution
in subsequent work -- for instance, if
it is desirable to determine the sensitivity
of the solution to changes in a parameter.

In contrast to the contour integral methods,
our geometric-analytic method is quite simple
to describe.  The only concept which may cause
a student some concern will be the use of the
Lambert W function, and we have tried to
motivate that in context.  It may be that
the student will find this solution of a
practical problem, the quantum square well,
provides a comprehensible introduction to
the Lambert W function as one of the family
of special functions which are useful not only
for quantum problems but for an immense
variety of problems in diverse fields.




\section{Solution using Lambert W}
\label{sect-solution}

This presentation of the solution of the
finite square well (FSW) problem will move fairly rapidly
through the parts of the solution process
which are familiar from standard textbooks,
and more gradually through the novel parts
of the solution.  
We will write the solution in terms of
complex variables as long as feasible,
rather than introducing real and imaginary
components too early.

Suppose a bound particle in a 1-dimensional
finite square well potential $V(x)$, 
which is given by
$V(x) = 0$ for $x < -L$ or $L < x$,
and by $V(x) = -V_0$ for $-L < x < L$.
Here $L$ and $V_0$ are positive reals,
$L$ being (half) the width of the well,
and $V_0$ being the depth.
The particle has energy $-E$, 
where $E$ is a positive real
between 0 and $V_0$.
The particle's wave function $\psi(x)$ satisfies
the time-independent Schr\"odinger
equation (TISE),
\begin{equation}
  \Big(-{\frac{\hbar^2}{2m}}\Big) \,\,
   \psi^{''}(x) \,\, + V(x)\,\psi(x) = -E\,\psi(x)
\end{equation}
There are three regions:
region 1 is $x < -L$, with wave function $\psi_1$;
region 2 is $-L < x < L$, with wave function $\psi_2$;
region 3 is $L < x$, with wave function $\psi_3$.

In region 2, the TISE becomes
\begin{equation}
  \psi^{''}(x) + \alpha^2\,\psi(x) = 0
\end{equation}
where $\alpha^2 = {(2m/\hbar^2)}\,(V_0-E)$
and the wave function satisfies
the equation
\begin{equation}
\label{eqpsi2}
  \psi_2(x) = A\,e^{-i \alpha x} + B\,e^{i \alpha x}
\end{equation}
for some complex constants $A$ and $B$, to be determined.
The value of $\alpha$ is real, and we can take it to be positive.

In regions 1 and 3, the TISE becomes
\begin{equation}
  \psi^{''}(x) - \beta^2\,\psi(x) = 0
\end{equation}
where $\beta^2 = {(2m/\hbar^2)}\,E$
The value of $\beta$ can be taken to be positive real.
The wave function satisfies
the equation
\begin{equation}
\label{eqpsi1}
  \psi_1(x) = D\,e^{\beta x}
\end{equation}
in region 1,
and the equation
\begin{equation}
\label{eqpsi3}
  \psi_3(x) = C\,e^{- \beta x}
\end{equation}
in region 3,
for some complex constants $C$ and $D$, to be determined.

The continuity constraints can be expressed as four equations:
\begin{align}
\label{eqcvpa}
  \psi_2(L)
     &= A \, e^{-i \alpha L} + B \, e^{i \alpha L} \\
     \nonumber &= C \, e^{- \beta L}
     = \psi_3(L)
\end{align}
\begin{align}
\label{eqcvna}
  \psi_2(-L)
     & = A \, e^{i \alpha L} + B \, e^{-i \alpha L} \\
     \nonumber & = D \, e^{- \beta L}
     = \psi_1(-L)
\end{align}
\begin{align}
\label{eqcdpa}
  \psi_2'(L)
     & = -i \alpha \, A \, e^{-i \alpha L} 
       + i \alpha \, B \, e^{i \alpha L} \\
    \nonumber & = - \beta \,C \, e^{- \beta L}
     = \psi_3'(L)
\end{align}
\begin{align}
\label{eqcdna}
  \psi_2'(-L)
     & = -i \alpha \, A \, e^{i \alpha L} 
       + i \alpha \, B \, e^{-i \alpha L} \\
     \nonumber & = \beta \, D \, e^{- \beta L}
     = \psi_1'(-L)
\end{align}

Form linear combinations of these equations as follows:
$i \alpha$ times equation (\ref{eqcvpa}) plus or minus
equation (\ref{eqcdpa}), and
$i \alpha$ times equation (\ref{eqcvna}) plus or minus
equation (\ref{eqcdna}).
Conceptually, those manipulations correspond to factoring
\begin{equation}
  - \big( \alpha^2 \psi + {\psi'}^2 \big) =
  \big( i \alpha \psi + \psi' \big) \, \big( i \alpha \psi - \psi' \big)
\end{equation}
and requiring that one of the right-hand factors be zero,
when evaluated at each of the points $x=-L$ and $x=L$.
We obtain four new equations:
\begin{equation}
\label{eqbc}
    2 i \alpha B \, e^{i \alpha L}
     = ( i \alpha - \beta ) \, C \, e^{- \beta L}
\end{equation}
\begin{equation}
\label{eqac}
    2 i \alpha A \, e^{- i \alpha L}
     = ( i \alpha + \beta ) \, C \, e^{- \beta L}
\end{equation}
\begin{equation}
\label{eqbd}
    2 i \alpha B \, e^{- i \alpha L}
     = ( i \alpha + \beta ) \, D \, e^{- \beta L}
\end{equation}
\begin{equation}
\label{eqad}
    2 i \alpha A \, e^{i \alpha L}
     = ( i \alpha - \beta ) \, D \, e^{- \beta L}
\end{equation}

One way to proceed is to divide equation (\ref{eqbc})
by equation (\ref{eqad}), and realize that $B/A$ and
$C/D$ are equal.  Let $\epsilon$ denote $B/A$.
Similarly, dividing equation (\ref{eqac}) by equation (\ref{eqbd})
shows that $A/B$ also equals $C/D$.  
Hence $\epsilon$ equals $1/\epsilon$, which leads
to two cases: $\epsilon$ is either 1 or -1.

That corresponds to the conclusion, in the textbook solution,
that $\psi$ must be either an even function or an odd function.
If $\epsilon$ is 1, then 
\begin{equation}
  \psi_2(x) = A(\,e^{-i \alpha x} + e^{i \alpha x}),
\end{equation}
which is the even function $\mathrm{cos}(\alpha x)$
times a complex constant $2A$, whose phase may be
chosen arbitrarily.
If $\epsilon$ is -1, then
\begin{equation}
  \psi_2(x) = A(\,e^{-i \alpha x} - e^{i \alpha x}),
\end{equation}
which is the odd function $\mathrm{sin}(\alpha x)$
times a complex constant $-2iA$, whose phase may be
chosen arbitrarily.

In the solution of
Bransden and Joachain \cite{Bransden-1989}, section 4.6,
which works with real values and hence
trigonometric functions, that dichotomy eventually results
in the two distinct equations (\ref{equtanv}) and (\ref{equcotv}).
Here, we continue with a complex valued approach
to the FSW problem to get the insights
that brings.  So we will simply for the time being
let $\epsilon$ denote a value
which, in the solution, will be either 1 or -1.
That is, $\epsilon$ represents either of the two
square roots of unity.

The four equations 
(\ref{eqbc}) thru (\ref{eqad}) 
reduce to two equations:
\begin{equation}
\label{eqac1}
    2 i \alpha \epsilon A \, e^{i \alpha L}
     = ( i \alpha - \beta ) \, C \, e^{- \beta L}
\end{equation}
\begin{equation}
\label{eqac2}
    2 i \alpha A \, e^{- i \alpha L}
     = ( i \alpha + \beta ) \, C \, e^{- \beta L}
\end{equation}
Dividing equation (\ref{eqac1}) by (\ref{eqac2}) gives
(since $\epsilon = \pm1$, $-\epsilon = \pm 1$)
\begin{equation}
\label{eqeps1}
  \epsilon \, e^{2 i \alpha L} =
  {\frac{\beta - i \alpha}{\beta + i \alpha}}
\end{equation}

Introduce variables $u = \beta L$ and $v = \alpha L$
to express (\ref{eqeps1}) as
\begin{equation}
\label{eqeps2}
  \epsilon \, e^{2 i v} 
  = {\frac{u - i v}{u + i v}}
  = {\frac{(u - i v)^2}{u^2+v^2}}
\end{equation}
The values of $u$ and $v$ are related to the energy $E$
via $u^2 = {(2m/\hbar^2)}\,EL^2$,
and $v^2 = {(2m/\hbar^2)}\,(V_0-E)L^2$,
so if we know $u$ or $v$, then the energy $E$
can be determined.
Moreover,
$u^2 + v^2 = {(2m/\hbar^2)}\,{V_0}L^2 = R^2$, say.
The values of $u$ and $v$ lie on an $R$-circle.
Here $R$ does not depend upon the energy, but is a
parameter of the FSW problem, depending only on the
(spatial) width and (energy) depth of the potential well.  
Bransden and Joachain call $R$ the ``strength parameter".
$R$ is unitless, and as will be seen,
the number of solutions of the FSW problem
will increase as the value of $R$ gets larger.

Equation (\ref{eqeps2}) thus simplifies to
\begin{equation}
\label{eqeps3}
  (u - i v)^2 = \epsilon R^2 e^{2 i v}
\end{equation}

Now we wish to introduce the Lambert W function.
Some references to Lambert W properties
and applications are
\cite{Corless-1996, NIST-2010, Valluri-2000}.  
For the present purpose, it suffices to know that
Lambert $W(z)$ is the analytic multi-branch solution
of $W(z) e^{W(z)} = w e^w = z$, where $z$ is the
complex argument of $W(z)$.
That is, if we can manipulate an equation into the
form $w e^w = z$, then the solution will be one 
or all of the branches of $w = W(z)$.

Comparing the Lambert function with the natural
logarithm function $\mathrm{log}(z)$, we observe that they are
closely related; $w = \mathrm{log}(z)$ is the multi-branch analytic
function which solves the equation $e^{\mathrm{log}(z)} = e^w = z$.
The natural logarithm is very familiar, and it has many useful
properties.  
Lambert W is similarly useful, once one learns to recognize 
problem situations where it has application.

Take the square root of equation (\ref{eqeps3}),
to obtain
\begin{equation}
\label{eqeps4}
  u - i v = \pm \sqrt{\epsilon} R e^{i v} = \gamma R e^{i v}
\end{equation}
Here we have let $\gamma$ represent the square root
of $\epsilon$, as well as a $\pm 1$ factor which
comes from taking the square root of $e^{2 i v}$.
That is, there are four alternatives;
$\gamma$ may be any of $\pm 1$ or $\pm i$.
Letting $\gamma$ represent an arbitrary member of
that set of four alternatives is convenient, since for
instance the conjugate, reciprocal or negative of
the symbol $\gamma$ is just $\gamma$, that is,
another representative of that set of four alternatives.
So we may take the conjugate in equation (\ref{eqeps4}),
replace the conjugate $\gamma^{*}$ by $\gamma$
(since $\gamma$ represents any one of the four complex
roots of unity),
and transfer the exponential to the left hand side,
to obtain a simpler-looking equation
\begin{equation}
\label{eqgam1}
  (u + i v) e^{iv} = \gamma R
\end{equation}

At this point we could, if we wished, 
replace $e^{iv}$ 
by $\mathrm{cos} \, v + i \, \mathrm{sin} \, v$,
and obtain
\begin{equation}
\label{eqgam2}
  (u + i v) (\mathrm{cos} \, v + i \, \mathrm{sin} \, v ) = \gamma R
\end{equation}
Suppose for instance that $\gamma = 1$.
Taking the imaginary part of equation (\ref{eqgam2}) gives
\begin{equation}
\label{eqimag1}
  v \, \mathrm{cos} \, v + u \, \mathrm{sin} \, v = 0
\end{equation}
That  equation (\ref{eqimag1}) is simply
\begin{equation*}
  v \, \mathrm{cot} \, v = -u
\end{equation*}
which, together with the constraint that
$u$ and $v$ lie on the circle of radius $R$, 
is the textbook version of the FSW solution.
However, we would like to keep the solution
process general for a bit longer.

We are almost there.  Consider the geometric
content of equation (\ref{eqgam1}).
Write $w = u + i v$, which, since we have
chosen $u$ and $v$ to be positive, lies in
the first quadrant.  Further, we know that
the magnitude of $w$ is $R$, since
$u^2 + v^2 = R^2$.
The effect of the factor $e^{i v}$ 
in equation (\ref{eqgam1}) is to rotate
$w$ counterclockwise by an angle of at most $R$ radians,
so that the product $w e^{i v}$ can lie in any
quadrant.  However, the condition that 
$w e^{i v}$ equal $\gamma$ (one of the four
fourth roots of unity) times the real number $R$,
means that only certain rotational angles $v$
are allowable as solutions of equation (\ref{eqgam1}).
The point $w e^{iv}$ cannot lie ``within" a
quadrant; it must lie on either the real or the
imaginary axis.
That of course is the resonance phenomenon
which is familiar with regard to stabilizing
the values of quantum mechanical observables.

Now, can we relate this to the Lambert W function?
Suppose we consider the real and imaginary parts
of the expression $z = w e^w$ where $w = u + i v$.
We have
\begin{equation*}
  z = w e^w = (u + i v) \, e^{u + i v} = e^u \, (u + i v) \, e^{i v}
\end{equation*}
\begin{equation*}
  = e^u \, (u + i v) \, (\mathrm{cos} \, v + i \, \mathrm{sin} \, v)
\end{equation*}
\begin{equation*}
  = e^u \, \big( u \, \mathrm{cos} \, v - v \, \mathrm{sin} \, v \big)
  + i \, e^u \, \big( u \, \mathrm{sin} \, v + v \, \mathrm{cos} \, v \big)
\end{equation*}

Imagine the mappings between the $z$-plane and the
$w$-plane.  The map $z = we^w$ carries the $w$-plane to
the $z$-plane, and the inverse map (multi-branch) is the
Lambert W function, carrying the $z$-plane to the $w$-plane.
The two rays from the origin in the $z$-plane along the
imaginary axis, are the values of $\gamma r$ when 
$\gamma = \pm i$ and $r$ is a positive real.
Those rays in the $z$-plane correspond to 
the $w$-plane values for which 
\begin{equation*}
 u \, \mathrm{cos} \, v - v \, \mathrm{sin} \, v = 0
\end{equation*}
which is equivalent to the equation 
\begin{equation*}
 v \, \mathrm{tan} \, v = u
\end{equation*}

Similarly, the two rays from the origin in the $z$-plane
along the real axis, are the values of $\gamma r$ when 
$\gamma = \pm 1$ and $r$ is a positive real. 
Those rays in the $z$-plane corresponds to the
$w$-values for which
\begin{equation*}
 u \, \mathrm{sin} \, v + v \, \mathrm{cos} \, v = 0
\end{equation*}
which is equivalent to the equation 
\begin{equation*}
 v \, \mathrm{cot} \, v = - u
\end{equation*}

Finally, the $R$-circle in the $w$-plane,
under the mapping $z = we^w$, has its image
as a closed (multi-loop self-intersecting) curve in the $z$-plane.

Those set correspondences show how to visualize the
solution of the FSW problem in terms of the Lambert W
function.  There are two alternative approaches:

(A)  Start with the axes (both real and imaginary) of the $z$-plane
excluding the point at the origin.
Let's call those two axes the sets $X$ and $Y$,
and let their union be $S = X \cup Y$.
That is, the set $S$ is four axial rays from the origin
in the $z$-plane.
Map the axial rays to the $w$-plane via the multi-branch
Lambert W function, obtaining the set $W(S) = W(X) \cup W(Y)$
as a family of lines in the $w$-plane -- one line for
each combination of an axial ray and a branch of the
Lambert W function.
Intersect that set $W(S)$, in the $w$-plane, with the circle 
$|w| = R$.  That is the solution set of the FSW problem.\\
or

(B)  Start with the circle $|w| = R$ in the $w$-plane,
let's call it $P$.  Map that circle to the $z$-plane via
$z = we^w$, to obtain a set, let's say $Q$ in the $z$-plane.
In a set-wise notation, we might write $Q = P \, e^P$.
Intersect that set with the axes $X$ and $Y$ in the $z$-plane.
That also is the solution of the FSW problem, with the
solution set represented in the $z$-plane.

It is useful to see these two approaches in graphical form.

\begin{figure}
\includegraphics[scale=1.0]
   {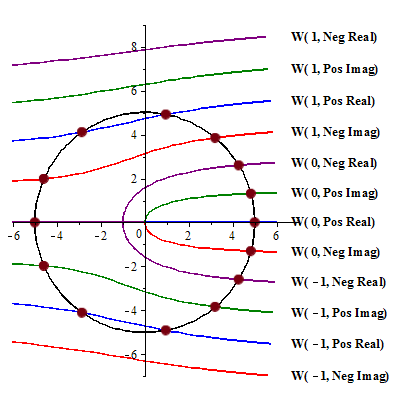}
\caption{\label{fig1}Solution in w-plane.
  $W(k,$ray$)$ means branch $k$ image of axial ray.}
\end{figure}

As an illustration, suppose that $R = 5$.
We will start with approach A, and observe that it
reproduces the textbook solution.
Figure 1 shows the $w$-plane.
The circle $|w|=5$ is shown, along with maps of the 
real and imaginary axes as transformed by the Lambert
W function.  Various branches of the Lambert W function
have been shown, for branch numbers 1 to -1 (using the
conventional definitions of branch numbers from Corless, et al
\cite{Corless-1996}).  Only the branch numbers 1, 0 and -1
intersect the circle $|w| = 5$.
The first quadrant of figure 1, 
for $u>0$ and $v>0$,
is readily seen to be equivalent to the textbook
diagram, with the $u$ and $v$ axes flipped.
Compare for instance figure 4.11 of
Bransden and Joachain \cite{Bransden-1989}; 
the authors, in their figure, use symbols $\xi$ for $v$,
$\eta$ for $u$, and $\gamma$ for $R$.
Their $\gamma=5$ curve shows the four
bound state solutions, two even states and
two odd states, as visible in the first quadrant of
our figure 1.

Now, what happens if instead we use approach B
to look at the solution for the $R=5$ example,
working in the $z$-plane?  
Instead of the circle $P: \, |w| = 5$ in the $w$-plane
we have its transform $Q$ under the map
$z = w e^{w}$, which gives a closed curve
in the $z$-plane, with multiple loops around
the origin.  We are interested in the places
where that curve crosses one of the axes,
since those are the possible solutions of the
finite square well problem.  Figures 2 to 5 
show that curve, 
the whole of the $Q$ curve in figure 2,
with magnifications in figures 3 to 5 to
exhibit the fine details.

\begin{figure}
\includegraphics[scale=0.5]
   {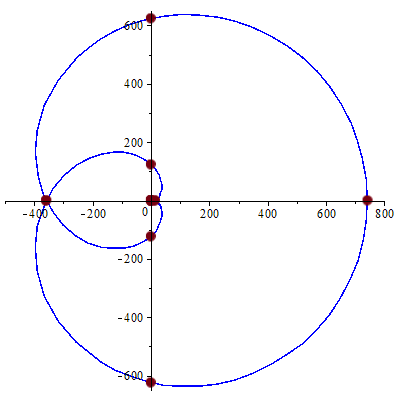}
\caption{\label{fig2}Solution in $z$-plane.}
\end{figure}

\begin{figure}
\includegraphics[scale=0.5]
   {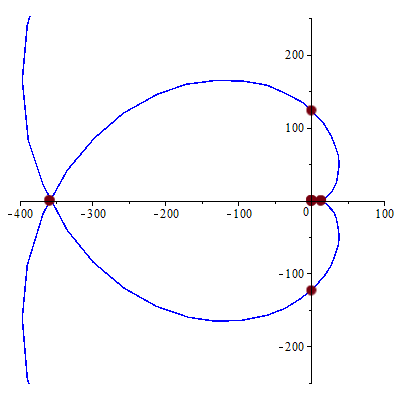}
\caption{\label{fig3}Solution in $z$-plane, magnified.}
\end{figure}

\begin{figure}
\includegraphics[scale=0.5]
   {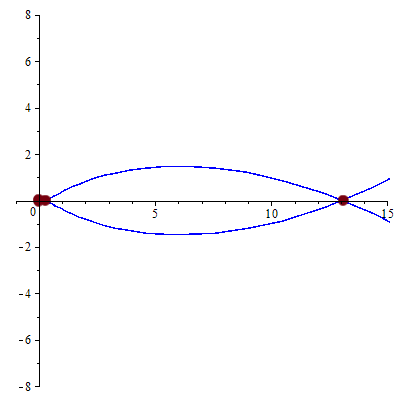}
\caption{\label{fig4}Solution in $z$-plane, magnified.}
\end{figure}

\begin{figure}
\includegraphics[scale=0.5]
   {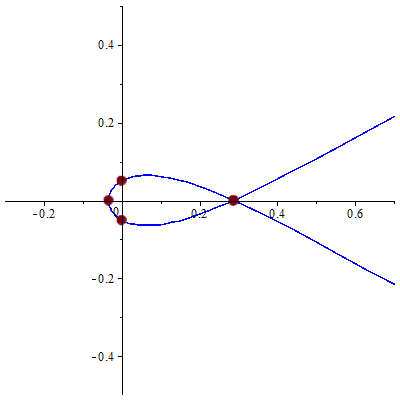}
\caption{\label{fig5}Solution in $z$-plane, magnified.}
\end{figure}

The $Q$ curve crosses the axes 14 times, counting
multiplicities, ie crossings at the same coordinates
but with different trajectories.  Those correspond to
the 14 solutions, the circle intersections in figure 1, of
the problem visualized in the $w$-plane.  To express
those solutions in terms of the phase $\theta$
of the $w$ variable, write 
$w = R \, \mathrm{exp}(i \theta)$.
Then the solutions (the phase angle $\theta$ parameter
values when the curve $Q$ crosses one of the axes) are: 
x-axis crossings at $\theta =$ 0.000, 0.546, 1.377, 2.179, 
3.142, 4.105, 4.906, 5.737, 
and y-axis crossings at $\theta =$ 0.264, 0.875,
2.735, 3.548, 5.408, 6.019.

To obtain the solutions in the first quadrant,
that is with $u$ and $v$ required to be positive, we can
simply restrict the $\theta$ phase to be between 0 and
$\pi/2$, and exclude the case $\theta=0$ as not being
physical.  That produces the standard textbook results.

As one can see the Lambert W function, considered as
a mapping between two planes, provides a visualization
of the solution of the finite square well problem in terms
of simple geometric objects: the real and imaginary axes,
and circles around the origin.  The area of the circle
around the origin corresponds to the dimensions of the
finite square well, and is proportional to the square of the
strength parameter, that is, to the depth of the
well and to the square of its linear dimension; in
symbols, $R^2 \sim V_0 L^2$.
The proportionality constant is $2m/{\hbar}^2$, and
the units of that constant cancel the units of $V_0 L^2$,
resulting in $R^2$, and hence $R$, being unitless.




\section{Discussion}
\label{sect-discuss}

The Lambert W description of the finite square well problem
is best visualized as a conformal map between two complex planes,
produced by the mapping $w \to z = w \, e^w$, the Lambert W
function being the multi-branch inverse of that mapping.

The axial rays in the $z$-plane map to the Lambert W lines
in the $w$-plane, whose intersections with the circle of
radius $R$ about the origin give the solutions to the FSW
problem.  Alternatively, the circle of radius $R$ about the 
origin in the $w$-plane maps to a multi-loop closed curve
in the $z$-plane, whose intersections with the axial rays
also give the solutions to the FSW problem.  Thus one may
approach the problem situation working in either the $w$-plane
(as is traditional), or in the $z$-plane -- the choice of
plane being determined by the convenience of other aspects
of the particular problem which may be simpler in one or
the other of those representations.

This technique bears some similarities to the method used
in \cite{Valluri-2000} to determine the fringing fields of
a parallel plate capacitor.

Because the mapping is conformal, the angles between the
circle in figure 1 and the various Lambert W lines are
equal to the angles of the corresponding intersections
in figure 2 of the multi-loop image of the circle and
the axial rays.  That suggests some possibilities for
design of materials to be sensitive to slight changes
in their environment, and leads back to the topic of the
quantum well infrared photodetector (QWIP) with which
we introduced this paper.

It is worth noting that the finite square well is ``realistic"
despite its simplicity, and continues to find use in 
contemporary research.
For instance, Deshmukh, et al \cite{Deshmukh-2014}
use a 3-dimensional radial finite square well model 
to characterize the attosecond-scale time delays of 
the photoionization of an atom of Xenon trapped 
within a C60 fullerine molecule.  
Kocabas, et al \cite{Kocabas-2009}, in a study
of mathematical models for metal-insulator-metal
waveguides, note (pages 13-14 of their paper) that
there is a close relationship between the one-dimensional
Schr\"odinger equation and the electromagnetic wave
equation in layered media, and mention several ideas
for investigation.  They write ``It is intriguing
to ask whether such studies [of FSW solutions and
their relationship to changes in reflection spectra
of wells] could be useful in optics for the 
investigation of the effects of material interfaces".
Thus there is plenty of opportunity to do interesting
work even with as old and familiar a topic as the
finite square well.

There is a relationship of the solutions of the finite
square well to the quadratix of Hippias, which is the
solutions in the $(x,y)$ plane of
the equation $x = y \, \mathrm{cot} \, y$.
You can flip the axes or make translations
to get other similar expressions of the
relationship between the two variables.
See Corless, et al \cite{Corless-1996}, page 344.

The quadratix of Hippias can be used to solve
various problems, such as the trisection of an
arbitrary angle.  Harper and Driskell \cite{Harper-2010}
have an enjoyable description of construction of the
quadratix, using interactive software for geometric
constructions, and show how to use the quadratix
to multiply an angle by any factor which can be
expressed as a ratio of the lengths of two straight
lines.  That raises the entertaining possibility that
one could perhaps present quantum mechanics
in the language of Euclidean geometry.  Physics
textbooks could return to the geometric methods 
utilized by Sir Isaac Newton in his \textit{Principia}!
That is a somewhat lighthearted suggestion, essentially
an entertainment, but it also represents a serious
alternative view -- and we may gain new insights.
Quantum mechanics is already often described with
a visual, diagrammatic language.

We anticipate there may be other interesting aspects of this
geometric-analytic solution technique, for future exploration.


\section{Conclusion}
\label{sect-conclusion}

We have presented a solution of the
quantum mechanics problem of the
allowable energy levels of a bound
particle in a 1-dimensional 
finite square well potential.
The solution is a ``geometric-analytic" technique
utilizing the Lambert W function.
The solutions can be represented in
either of two domains, and a representation
in the transformed domain has a particularly
simple geometry.
There are many opportunities to follow up.
More work on this vibrant field of the application
of the Lambert W function is anticipated.

SRV thanks King's University College for
their generous support.





\end{document}